\documentstyle[aps,prl,psfig]{revtex}
\newcommand{\sngl}[2]{#1$_{#2/2}$}

\title{Wobbling motion coupled to  gamma vibration at high spin}
\author{Makito Oi$^{1,2}$\footnote{m.oi@surrey.ac.uk} and Philip M. Walker$^1$}
\address{$^1$ Department of Physics, University of Surrey, 
Guildford, Surrey, GU2 7XH, United Kingdom.\\
$^2$ Department of Applied Physics, Fukui University, 3-9-1 Bunkyo,
Fukui 910-8507,Japan.}
\date{\today}

\begin{document}
%\twocolumn[
%\columnwidth\textwidth\csname@twocolumnfalse\endcsname
\maketitle
\begin{abstract}
We report a solution of the tilted-axis cranked HFB equation
for $^{164}$Hf,  which shows wobbling motion coupled to gamma
vibration at high spin ($J\simeq 60\hbar$).
Possible anharmonicity and splitting of energy levels are also discussed as
a consequence of the wobbling motion with large amplitude.

%\pacs{
PACS number(s): 27.70.+q, 21.10.-k
%}
\end{abstract}

\addvspace{7mm}
%]
%%%%%%%%%%%%%%%%%%%%%%%%%%%%%%%%
% body of paper here
%%%%%%%%%%%%%%%%%%%%%%%%%%%%%%%%
%
In understanding the excited states
of a quantum many-body system with finite degrees of freedom,
the concept of dynamic collective modes has been playing a leading role.
For example, in nuclear physics,
shape vibrations such as $\gamma$ and $\beta$ vibrations
\cite{BM75}, octupole vibrations in super-deformed states\cite{NMM93}, 
and giant dipole resonances \cite{RS80} are well known.
In a dilute atomic gas 
in a Bose-Einstein condensate, which is also a finite system,
 the dynamic motion displays the scissors mode, which is an out-of-phase
vibration among states in the super and normal fluidity \cite{BEC}.

As well as these vibrations, rotations have been also considered
and studied in detail. As a result, rotational bands are observed
in nuclei, as in atomic and molecular systems \cite{Ra50,BM75}.
The relevant rotational forms are essentially uniform and one-dimensional,
and they are sometimes called "static rotation" because the rotation axis is 
fixed.

Advances in experimental techniques these days allow us
to obtain data for excited levels  built on very high spin states.
For understanding these levels,
a concept of ``dynamic'' nuclear rotation is introduced and 
recently has attracted much interest.
Indeed, the concept is helpful to comprehend
several novel phenomena, such as
nuclear wobbling motions in triaxial deformed nuclei 
\cite{BM75,Ode01}, 
dynamical coupling modes (wobbling modes) between high- and low-$K$ bands
\cite{OA00},  and
chiral vibrations through quantum tunnelling 
between right- and left-handed chiral rotating states \cite{SK01}.
These phenomena have recently been studied 
from both experimental and theoretical approaches.

Dynamic rotation is usually considered in terms of the  classical mechanics 
as time-dependent and three-dimensional rotation,
such as the precession in the rotation of the earth \cite{Gold},
 or motions of a tippe top in which the direction 
of the rotation axis varies time-dependently 
in the intrinsic frame \cite{ARD55}.
The discussion of these rotations in nuclear physics was
first presented 
by Bohr and Mottelson in the context of wobbling motion 
\cite{BM75}.
They considered the motion 
as an analogy of the small orientation fluctuations of 
a triaxially deformed classical rigid rotor from its
static and one-dimensional rotation about the principal axis
of the inertia tensor.
They quantize the Hamiltonian of the rotor 
under the condition $|I_1|\simeq |I| >>1$,
by using the technique similar to the second quantization 
of the harmonic oscillator. ($I$ is the total angular momentum and 
$I_1$ is the $x$ component of the angular momentum vector.)
It is then shown  that the excited levels
corresponding to the nuclear wobbling motion  have
a vibrational character on the rotational band, 
$E(I,n_{\rm w})=\frac{I(I+1)}{2{\cal J}_1}+\hbar\omega_{\rm
w}\left(n_{\rm w}+1/2\right)$,  where $\hbar\omega_{\rm w}$ is calculated by
the moments of inertia of the rotor and ${\cal J}_1$ is 
the moment of inertia around the $x$-axis. The oscillation quanta 
$n_{\rm w}$
take the values $n_{\rm w}=0,1,2,\cdots$.
In spite of their theoretical prediction, it had been difficult
to identify the wobbling states in experiments until 
the experimental evidence 
recently reported in $^{163}$Lu with triaxial super-deformation
\cite{Ode01}. 
With this discovery, 
there has been established a new research field in high-spin nuclear physics.

Nuclear rotation is closely related to the shape of a nucleus or its symmetries
\cite{BM75}, so that exotic rotations can be seen for exotic shapes.
Nuclei possessing  axial symmetry follow the
static one-dimensional rotation, 
such as in the ground-state rotational band, 
or g-band. 
These rotations are  low-lying collective modes which are caused
by the spontaneous symmetry breaking mechanism for the rotational
symmetry, that is, the SU(2) symmetry is broken 
into the subgroup U(1) through the nuclear deformation.
As the  U(1) symmetry is further broken into its discrete 
sub-symmetries,
more variations can be expected for the nuclear rotation.
For instance, 
dynamic and three- (or two-) dimensional rotations are predicted
to emerge in nuclei with triaxial symmetry 
(a discrete symmetry for a $\pi$ rotation around any axis, 
denoted as $D_2$).

One theoretical method  used for the analysis of the wobbling motion
is based on the particle-rotor model (PRM) \cite{Ode01,HM83,Ham87}.
Although the model allows quantum mechanical studies,
the assumption of a ``rotor'' brings a macroscopic treatment into the
analysis.
The microscopic formulations for the dynamic rotations
were derived in the framework of
 the random phase approximation (RPA) \cite{JM79}, and
 the time-dependent variational method \cite{Ma79,KO81}. 
Based on these and all subsequent works \cite{HO96,DZR99,OA00},
 we are now able to study 
the dynamic nuclear rotations from a microscopic point of view.
The combination of 
the tilted-axis cranking model \cite{HO96,Fr00,Bng93,MMM00} 
and  the generator coordinate method (GCM) \cite{FV75,RS80,OA00} 
is one of the methods which are very powerful and useful in
numerical sutdies.

Despite the fact that the experimental
discovery of the wobbling mode was made in an odd-mass nucleus ($^{163}$Lu)
and the theoretical investigation that the wobbling motion is favoured
in odd-$A$ mass nuclei \cite{Ham87}, 
it is essentially 
the ``rotor'' (in the PRM)
that wobbles. Hence, it is necessary
to study the wobbling motion in even-even nuclei ( i.e., nuclei
consisting of the even numbers of protons and neutrons,
where all the nucleons are coupled in the ground state 
through the BCS-type pairing interaction).
According to Ref.\cite{Bng01}, it is calculated that
the yrast state (the lowest energy state
for a given spin) in $^{164} _{72}$Hf becomes triaxial super-deformed (TSD)
at $I\simeq 40\hbar$, having gamma deformation $\gamma \simeq -20^{\circ}$. 
(Our convention for
the gamma deformation parameter,$\gamma$, is defined as in p.7 of
Ref.\cite{RS80}.)
Thus, $^{164}$Hf is expected to 
show the wobbling mode as a result of its triaxially deformed states
at high spin.

In this paper, we thus investigate the nuclear wobbling motion for 
$^{164}$Hf,
by means of self-consistent calculations
based on the tilted-axis cranking model.
However, before going forward to the microscopic analysis,
it is worth reviewing rotation
of a classical rigid body 
because the analogy with  classical mechanics is quite helpful
to understand nuclear wobbling motions.

\begin{figure}[h]
\begin{center}
\leavevmode
\parbox{0.9\textwidth}
{\psfig{figure=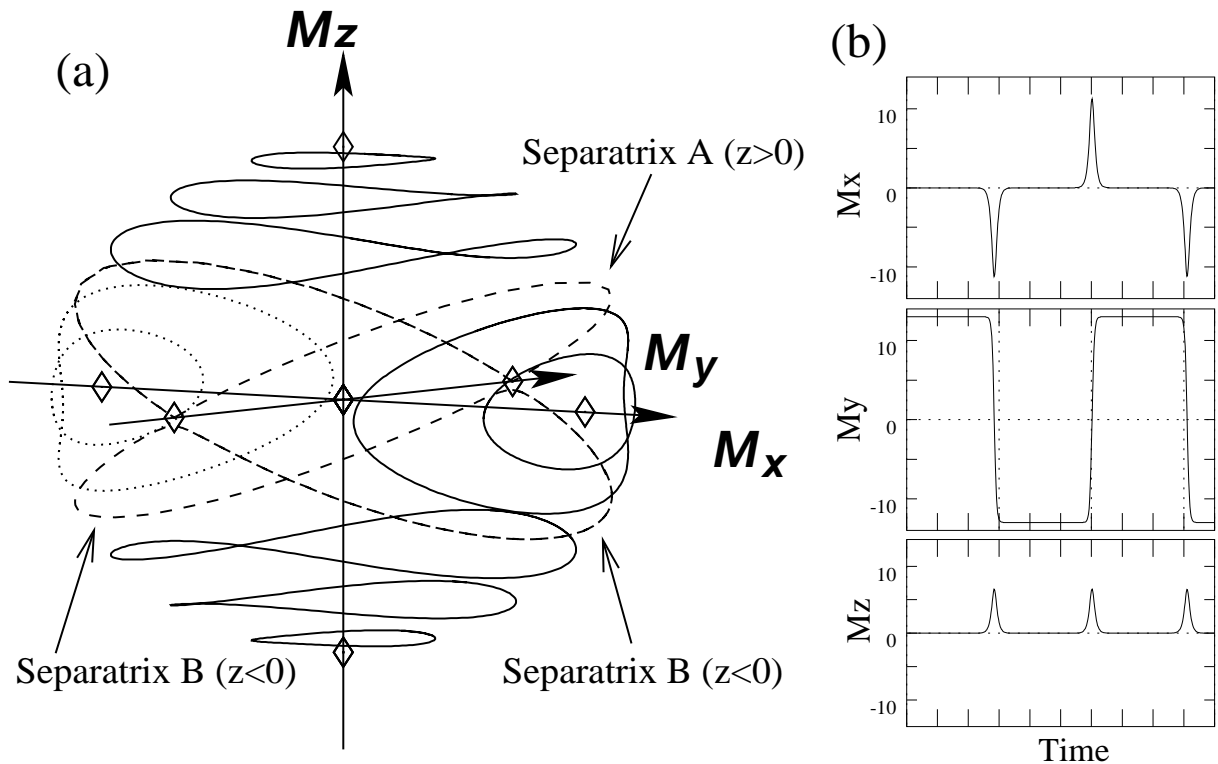,width=0.9\textwidth}}
%
%\parbox{0.5\textwidth}
%{\psfig{figure=fig1.eps,width=0.45\textwidth}}
\end{center}
%\mediumtext
\caption{(a): Trajectories of the angular momentum vector of a classical
rigid body (with a quadrupole shape;
 $(\beta,\gamma)=(0.4,-20^{\circ}$)). 
Separatrices are denoted by chains.
(b): Time dependence of angular momentum components
 on the separatrix  in the region $z>0$ 
(written as ``Separatrix A ($z>0$)'' in the figure). 
}
\label{fig1}
\end{figure}

Rotations of a classical rigid body can be studied through
the corresponding angular momentum vector 
\cite{Gold,HM83,Ons83}.
If the intrinsic coordinates are chosen so as to diagonalize the inertia
tensor, 
cross sections between a sphere,
$\displaystyle
 M^2 = M_1^2+M_2^2+M_3^2$ (angular momentum conservation), 
and an ellipsoid,
$ E = \frac{M_1^2}{2{\cal J}_1} 
 + \frac{M_2^2}{2{\cal J}_2} 
 + \frac{M_3^2}{2{\cal J}_3}$ (energy conservation),
give the trajectory of the vector on the sphere.
($M_i$ denotes the $i$-th component of the ``classical'' angular
momentum.)
Fig.\ref{fig1}(a) shows examples of such trajectories.
When the rigid body has  axial symmetry (say, along the
$z$-axis), it is well known that 
the trajectories follow ``precession'', which means that
motions of the vector are circular around the $z$-axis
and in the plane parallel to the $x-y$ plane.
However, in the case of a triaxial rigid body (${\cal J}_1 \ne {\cal
J}_2 \ne {\cal J}_3$), the trajectories deviate from the precession
in two ways: (i) distortion from the circle and (ii) deviation from
two-dimensional to three-dimensional movement.
These deviations imply that additional vibrations
are induced 
by breaking the axial symmetry of the rigid body.
These vibrational motions in rotations 
are called ``wobbling motions'' \cite{Ons83}.
It should be noted that these deviations from the precession are small
when the angular momentum is polarized near the axes ($x$- or $z$-axes
in Fig.\ref{fig1}). 
This type of small fluctuation corresponds to the nuclear wobbling
motion  in Ref.\cite{BM75} if the total angular momentum $|M|$ is large enough 
;$|M|\simeq |M_1| >> 1 $.
Dashed lines in Fig.\ref{fig1}(a) correspond to special trajectories
called ``separatrices'', which divide 
the area into four topologically different domains.
The existence of such domains 
is  evidence of non-linearity in the system. 
In fact, in Fig.\ref{fig1}(a), 
there are two separatrices separated at points 
${\cal P} \ (M_1,M_2,M_3)=(0,\pm 1,0)$,
and  they do not cross each other, as  shown in Fig.\ref{fig1}(b).
This subset figure shows the time-dependence of the angular momentum
vector for the separatrix in the domain $z>0$, which is obtained
by solving the Euler equations ($\frac{d\omega_i}{dt} = \frac{{\cal J}_j
 - {\cal J}_k}{{\cal J}_i}  \omega_j\omega_k,$ )
with proper initial conditions. 
(Indices $(i,j,k)$ should be taken in a cyclic manner).
Most of the time, the vector stays 
at $\cal P$ and 
the motion between the two points is comparatively fast
once the vector is apart from these points.
 The other separatrix has  similar motion 
except the sign of $M_z$.

Now, let us begin the microscopic analysis by means of the
tilted-axis cranked HFB method. 
Our Hamiltonian consists of two terms: a spherical part (
the spherical Nilsson Hamiltonian) and a residual part
(pairing-plus-$Q\cdot Q$ force).
The model space and  force parameters are chosen in the same manner as 
in our previous work \cite{OWA01}.
The HFB equation is solved by means of the method of steepest descent,
under constraints on the angular momentum and the particle number.
Particularly, the angular momentum constraints are expressed as,
$
 \langle \hat{J}_1\rangle = J \cos\varphi, \
 \langle \hat{J}_2\rangle = 0, \
 \langle \hat{J}_3\rangle = J \sin\varphi,
$
where $\varphi$ denotes the tilt angle.
For details about the method, see Ref.\cite{HO96}.
An advantage of the present method is that the physical quantities like
the deformation parameters are calculated self-consistently.
It is thus possible to describe the deformations as functions
of the tilt angle, for instance, $\gamma=\gamma(\varphi)$.

First of all, let us see briefly the results obtained by the
principal-axis (one-dimensional) cranking calculations ($\varphi=0^{\circ}$).
In the present calculations, 
there are no super-deformed solutions appearing in the yrast band 
throughout the range $0\le J \alt 80\hbar$. This can be explained
by that fact 
that our model space is not large enough to reproduce super-deformation.
Nevertheless, 
at $J\simeq 60\hbar$, the yrast states take on gamma deformation 
($\gamma\simeq -20^{\circ}, \beta\simeq 0.13$). 
For the purpose of studying the wobbling motion theoretically,
emergence of triaxility is enough.
The  gamma deformed states appear after
of the gap energy vanishes at $J\simeq 50\hbar$ $(20\hbar)$ 
for protons (neutrons). In the present calculations, major 
contributions to the gamma deformation
are alignments of \sngl{i}{13} and \sngl{h}{11}  protons 
in addition to neutron alignments at lower spins ($J\simeq 20\hbar$);
\sngl{i}{13}, \sngl{f}{7}, and \sngl{h}{9}.

Next, let us see the results in the tilted-axis (two-dimensional)
cranking calculations. 
The tilted-axis cranked HFB states are created from the 
principal-axis cranked HFB states $(\varphi=0^{\circ})$ at $J=60\hbar$
as the initial states.
The range of the gamma deformation is 
self-consistently determined  and given in the range
$180^{\circ}\alt \gamma \alt 300^{\circ}$.
Due to the symmetry in the definition of  $\gamma$, the range 
may be converted into  $-60^{\circ}\le \gamma \le 60^{\circ}$. 
The conversion is equivalent to renaming the axes in a cyclic 
manner, for instance, ``$x$-axis $\rightarrow$ $y$-axis''.
As a result, the wobbling motion in this study is in the $x$-$y$ plane.
In other words,
 it can be said that the tilt angle in the present work 
 corresponds to the azimuthal angle
 ($\varphi$) while  in the previous works \cite{OA00,OWA01} 
it corresponds to the polar angle ($\theta$), if the $z$-axis is chosen for the
direction of elongation of the nucleus.

\begin{figure}[h]
 \begin{center}
  \leavevmode
  \parbox{0.95\textwidth}
  {\psfig{figure=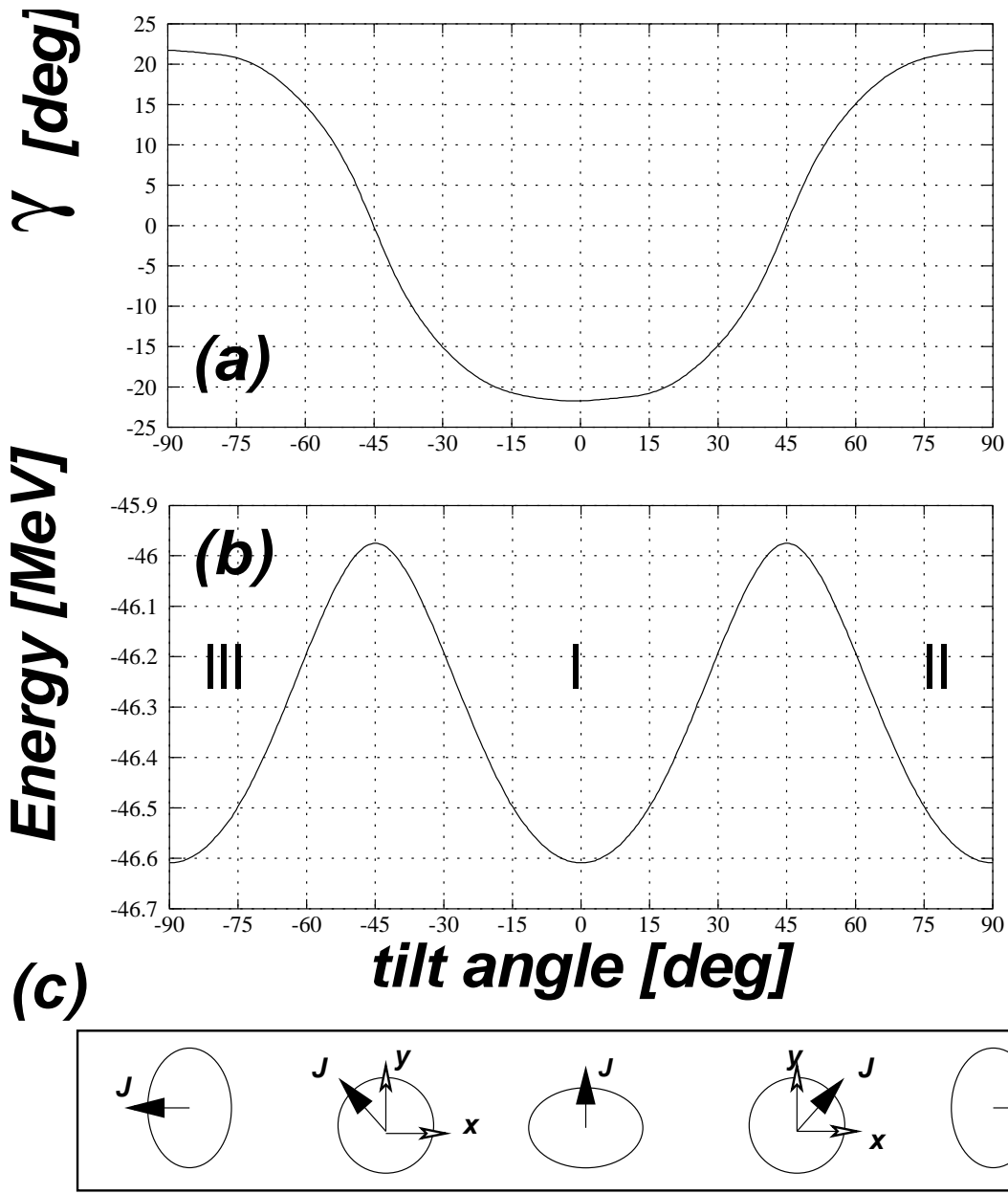}}
%
%\parbox{0.5\textwidth}
%{\psfig{figure=fig2.eps,width=0.45\textwidth}}
 \end{center}
%\mediumtext
\caption{
(a) Gamma deformation with respect to tilt angle, $\gamma(\varphi)$;
(b) Energy curve with respect to tilt angle, $E(\varphi)$; and
(c) a schematic picture of wobbling motion coupled with gamma vibration.
At $\varphi=\pm 45^{\circ}$ the nucleus has the axial symmetry, and 
the state corresponds to the separatrix in  classical mechanics.
}
\label{fig2}
\end{figure}

Figs.\ref{fig2}(a) and (b) show, respectively, changes of
gamma deformation,$\gamma(\varphi)$,  and energy, $E(\varphi)$.
From (a), coupling of the tilting and gamma degrees of 
freedom is clearly seen.
It should be noted that the curvatures of $\gamma(\varphi)$
are flat around $\varphi\simeq 0^{\circ}$ and $\pm 90^{\circ}$.
This result implies that the assumption in the PRM studies (fixing
the gamma values) \cite{Ode01,Ham87}
is reasonable as long as the wobbling amplitude is small.
For large wobbling amplitude, it seems that
the $\gamma$ vibration is cooperatively induced by the fluctuation of
the orientation of the rotation axis, that is, the wobbling motion.
This situation is schematically depicted in Fig.\ref{fig2}(c) .
(The $\beta$ values are almost constant, $0.130\alt \beta \alt
0.135$, for the entire range of $\varphi$).
The nuclear shape becomes prolate
when the tilt angle reaches $\varphi=\pm 45^{\circ}$, 
which correspond to unstable extrema in Fig.\ref{fig2}(b).
At these points, the $\varphi$ range is divided into several domains
from a dynamical point of view. 
Let us call the domains for $-45^{\circ} \le
\varphi < 45^{\circ}$ domain-I, in which the wobbling motion is around
the $y$-axis, while the other for $45^{\circ} <
\varphi \le 135^{\circ}$ domain-II, in which the wobbling is 
around the $x$-axis.
Domain-I and domain-II are related by a discrete symmetry operation
called $C_{4z}$ ( a symmetry for $\frac{\pi}{2}$-rotation about the $z$-axis),
due to the dynamical coupling between $\varphi$ and $\gamma$.
The wobbling motions are topologically classified through these domains.
Therefore, it can be said that 
the rotations at $\varphi=\pm 45^{\circ},\pm 135^{\circ}$ correspond to
``separatrices'' in  classical mechanics.
As seen in Fig.\ref{fig2}(b), the energy curve looks like a 
harmonic potential near $\varphi \simeq 0^{\circ}$ and
$\simeq90^{\circ}$. 
From the present calculation, it is obtained as
$V(\varphi)\simeq a\varphi^2$ where $a=1.64$ (MeV).
As $\varphi$ approaches the value for the
separatrix, the curve deviates from the harmonic shape.
Similar features are seen in the wobbling motion of a classical rigid body
with triaxiality, that is, the deviation from  planar motion
becomes larger as the trajectory reaches  a separatrix.

Let us discuss the possible manifestation of the excited structures
of the wobbling motion.
In Ref.\cite{BM75}, 
the wobbling excitations are
 given as a harmonic structure like $\hbar\omega_{\rm w}(n_{\rm
w}+1/2)$. 
Although the harmonic levels are expected to be seen for 
the wobbling motion with small wobbling amplitude, i.e., small $n_{\rm w}$,
anharmonicity can be expected
for wobbling motions with  large amplitude, i.e., large $n_{\rm w}$,
according to the shape of the energy curve.
To see the  excited  structure schematically,
we solve the one-dimensional Schr\" odinger equation for 
a wobbling phonon, i.e., the quantized wobbling motion.
It is assumed that the tilt angle can be treated as a dynamical variable,
and that 
$E(\varphi)$ corresponds to the potential $V(\varphi)$.
By putting walls with infinite height at $\varphi=\pm 45^{\circ}$,
the wobbling phonon is confined in the domain-I for the sake of
simplicity.
The formula (4-304) in Ref.\cite{BM75} for the wobbling phonon
energy ($\hbar\omega_{\rm w}$) and 
the coefficient of the approximated  harmonic potential, 
$a\equiv M_{\rm w}\omega_{\rm w}^2/2$, which
is obtained earlier, are used to obtain
the ``mass'' for the wobbling phonon,
 $M_{\rm w}/\hbar^2=a/(\hbar\omega_{\rm w})^2$.
The expression for the moment of inertia is employed from 
the expression  for a rigid body with 
quadrupole deformation ${\cal J}_i = \frac{2{\cal M}R^2}{5}\left( 1-
\sqrt{\frac{5}{4\pi}}\beta\cos\left(
\gamma-\frac{2\pi}{3}i\right)\right)$, in which
$\cal M$ is the mass of the nucleus ($1.53\times 10^5$ MeV \cite{MN})
and $R=1.2A^{1/3}$ (fm) with the mass number $A=164$.
The calculated wobbling phonon energy $\hbar\omega_{\rm w}$ is 
given to be 70 (keV)  where the quadrupole deformations
are $(\beta,\gamma)=(0.13,-20^{\circ})$ at 
$\varphi\simeq0^{\circ}$ and $J=60\hbar$.
The results are shown in Fig.3. 
Because $V(\varphi)$ keeps the 
harmonic shape fairly well 
up to $|\varphi|\alt 0.4$ $(i.e., 23^{\circ})$, 
it is seen that the quantized energies for the 
wobbling motion ($n_{\rm w} \alt 5$) 
possess the similar structure  to the harmonic case, 
i.e., equal energy spacing. 
On the other hand, for $n_{\rm w} \agt 5$,
the structure starts to deviate from the harmonic structure.

Unlike the classical wobbling motion of a triaxial rigid body,
tunnelling effects should be taken into account in the nuclear
wobbling motion. In other words, each wobbling mode classified by
domains can jump from one domain to another over the separatrices.
As a result of the coupling between $\varphi$ and $\gamma$ degrees of freedom, 
$E(\varphi)$ has discrete  $C_{4z}$ symmetry, implying 
that a wobbling phonon moves in the four potential wells.
It is well known that the double well potentials 
cause the splitting in energy levels due to the parity symmetry 
in the corresponding Hamiltonian \cite{Sa94}.
The states are then expressed by symmetric 
and antisymmetric linear combinations 
of the localized states in each potential.
In the present case, 
there would be four degenerate states localized in each potential,
if the barrier height were infinitely high. Let us denote these 
localized states as $|{\rm L1}\rangle$, $|{\rm L2}\rangle$, $|{\rm R2}\rangle$,
and $|{\rm R1}\rangle$, for the further left potential to the further
right. 
For the parity operation $\hat{\pi}$, ``R'' and ``L'' are exchanged.
For instance, $\hat{\pi}|{\rm L1}\rangle = |{\rm R1}\rangle$.
In the case of finite barrier height,
these four localized states are superposed through the tunnelling to produce 
two symmetric and two antisymmetric linear combinations.
These states might form a quartet, and be written as,
\begin{eqnarray}
|{\rm S1}\rangle &=& \frac{1}{2} 
 \left(
  |{\rm L1}\rangle  +  |{\rm L2}\rangle  +
  |{\rm R2}\rangle  +  |{\rm R1}\rangle 
\right)
\\
|{\rm S2}\rangle &=& \frac{1}{2} 
 \left(
  |{\rm L1}\rangle  -  |{\rm L2}\rangle  -
  |{\rm R2}\rangle  +  |{\rm R1}\rangle 
\right)
\\
|{\rm A1}\rangle &=& \frac{1}{2} 
 \left(
  |{\rm L1}\rangle  +  |{\rm L2}\rangle  -
  |{\rm R2}\rangle  -  |{\rm R1}\rangle 
\right)
\\
|{\rm A2}\rangle &=& \frac{1}{2} 
 \left(
  |{\rm L1}\rangle  -  |{\rm L2}\rangle  +
  |{\rm R2}\rangle  -  |{\rm R1}\rangle 
\right)
\end{eqnarray}
``S'' and ``A'' denote symmetric and antisymmetric states.
If each localized state is in the ground state, the nodal numbers for the
superposed states
are 0,1,2, and 3 for ``S1'', ``A1'', ``S2'' and
``A2'', respectively. The  order of 
the corresponding energy levels is thus expected to be 
 $E_{\rm S1} < E_{\rm A1} < E_{\rm S2} < E_{\rm A2}$.

In summary,
we have investigated the possible wobbling motion in $^{164}$Hf 
by means of the tilted-axis cranked HFB states. The calculations
show that the nucleus has sizeable gamma deformation at high spin
($J\simeq 60\hbar$), and wobbling motion coupled to the
gamma vibration can be expected. 
A possible anharmonicity in the wobbling excitations is predicted
for higher wobbling excitations,
based on the calculated energy curve with respect to the tilt angle.
In addition, a brief discussion is presented to show 
that a ``wobbling quartet'' can be expected for
the nuclear wobbling motion coupled to the gamma vibration at high spin,
as a consequence of the tunnelling effect.

M.O. thanks to Drs. I. Hamamoto, N. Tajima, T. D\o ssing, 
and P. Ring for fruitful discussions and hospitalities during his
stay in Lund University, Fukui University, Niels Bohr Institute,
and Technische Universit\" at M\" unchen.
Discussions with Drs. R. Bengtsson, G.B. Hagemann, A. Hayashi, A. Ansari, and
R. Hilton are also appreciated.
He would like to acknowledge financial support 
from the Japan Society for the Promotion of Sciences (JSPS), and
technical assistance from Center for Nuclear Sciences (CNS), University of
Tokyo, for parts of numerical calculations in this work.

%%%%%%%%%%%%%%%%%%%%%%%%%%%%%%%%
% end of body
%%%%%%%%%%%%%%%%%%%%%%%%%%%%%%%%

% now the references. delete or change fake bibitem. delete next three
%   lines and directly read in your .bbl file if you use bibtex.
%\vspace*{-6mm}

%%%%%%%%%%%%%%%%%%%%%%%%%%%%%%%
% Figures 
%%%%%%%%%%%%%%%%%%%%%%%%%%%%%%%%

\begin{figure}[h]

 \begin{center}
  \leavevmode
\parbox{0.4\textwidth}
{\psfig{figure=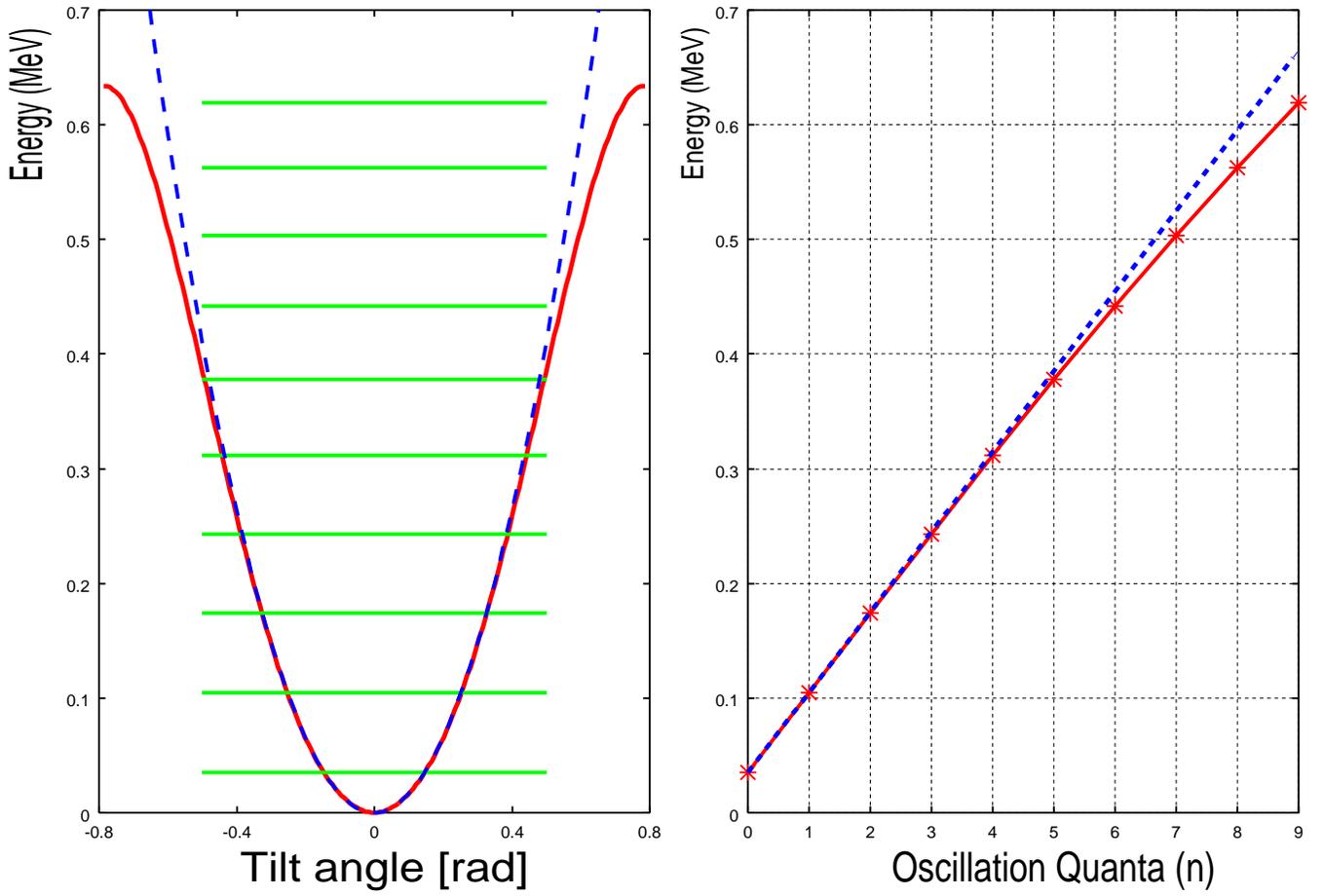,width=\textwidth,angle=-90}}
%
%\parbox{0.5\textwidth}
%{\psfig{figure=fig3.eps,width=0.45\textwidth,height=0.2\textwidth,angle=-90}}
\caption{Excited levels of the wobbling phonon.
Left: 
the solid curve is the energy (corresponding to Domain-I)
 calculated by the tilted-axis
cranked HFB method for $J=60\hbar$, while 
the dashed line is the harmonic approximation of the above line.
The range for the tilt angle
corresponds to  $|\varphi|\le 45^{\circ}$. Horizontal lines denote the 
quantized levels for the wobbling motion. Right: A simple line
indicates $\hbar\omega_{\rm w}(n_{\rm w}+1/2)$, while
the asterisks connected by lines 
show the quantized wobbling energies calculated 
in this study.
}

 \end{center}
\label{figaa}
\end{figure}

\end{document}